\documentclass[aps,prl,twocolumn,superscriptaddress,floatfix,amsmath,showpacs]{revtex4}

\usepackage{bm}
\usepackage{amsmath}
\usepackage{graphicx}

\begin{document}

\title{Quantum Condensation from a Tailored Exciton Population in a Microcavity}

\author{P. R. Eastham}
\affiliation{Department of Physics, Imperial College London, London SW7 2BW, U.K.}

\author{R. T. Phillips}
\affiliation{University of Cambridge, Cavendish Laboratory, Cambridge CB3 0HE, UK}

\date{\today}

\begin{abstract}
An experiment is proposed, on the coherent quantum dynamics of a
semiconductor microcavity containing quantum dots. Modeling the
experiment using a generalized Dicke model, we show that a tailored
excitation pulse can create an energy-dependent population of
excitons, which subsequently evolves to a quantum condensate of
excitons and photons. The population is created by a generalization of
adiabatic rapid passage, and then condenses due to a dynamical analog
of the BCS instability.
\end{abstract}


\maketitle

There is great interest in the possibility of quantum-condensed phases
of solid-state quasiparticles, such as excitons, polaritons, and
magnons.  Such phases are characterized by the presence of a quantum
state whose population scales with the size of the system, and hence
is much larger than one -- macroscopic occupation. This is seen in
recent experiments on Bose-Einstein condensation (BEC) of polaritons
and polariton lasing. In these experiments\
\cite{DPG+06,KRK+06,BGS+07,bajoni-2008,krizhan2}, a semiconductor
microcavity is excited at high energies, and a macroscopic population
of low-energy polaritons emerges following relaxation and inelastic
scattering\ \cite{DCT+06}. These condensates appear spontaneously,
from states without macroscopic occupations. This differentiates them
from the microcavity parametric oscillator experiments\
\cite{PhysRevLett.85.3680}, where resonant pumping of polaritons leads
directly to a macroscopic occupation.

The aim of this paper is to show how microcavities could be used to
access condensation, even in the absence of relaxation or inelastic
scattering. We propose an experiment on a microcavity containing an
ensemble of quantum dots, where the exciton decay times are many tens
or hundreds of picoseconds\ \cite{LP05}. We demonstrate that this
experiment could be faster than these decay times, so that energy
relaxation and inelastic scattering would be negligible. Nonetheless,
we shall show that a condensate develops. In contrast with a laser,
this condensate is formed from part-matter, part-light
quasiparticles. In contrast with the microcavity parametric
oscillator, it develops from a state with no macroscopic occupations,
in the absence of the pump laser. And whereas relaxation is essential
to obtain an equilibrium BEC or polariton laser, in our approach
condensation occurs due to an instability of the coherent quantum
dynamics. Our proposal implements, in a solid-state system, the type
of dynamical condensation predicted in quenched Fermi gases\
\cite{andreev04,barankov04,yuzbashyan06,eastham07}.

\begin{figure}[t]
\includegraphics[width=246pt]{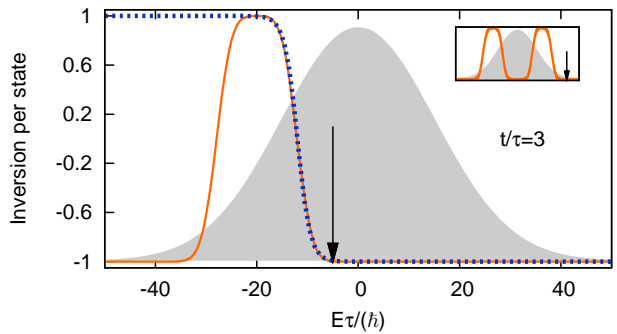}
\caption{(Color online) Simulated exciton inversion profile
  immediately after pumping, showing the population created by the
  Gaussian pump pulse (\ref{eq:gausspulse}) (solid line), and by a
  superposition of such pulses (inset).  The dotted curve is an
  equilibrium exciton distribution with fitted temperature $\hbar/(4.2
  k \tau)$; this is $0.6\mathrm{\ K}$ for $\tau=3\mathrm{\
  ps}$. Arrows mark the energy of the $\mathbf{k}=0$ cavity mode.
\label{fig:distn}}
\end{figure}

The first stage in our proposed experiment is the creation of a
population of excitons in the quantum dots. We propose using a chirped
laser pulse, which sweeps up through part of the
inhomogeneously-broadened exciton line. As shown by the demonstration
of Rabi oscillations\ \cite{ramsay:113302} and density-matrix
tomography\ \cite{WLD+06}, excitons in quantum dots are discrete
two-level systems, which can therefore be manipulated using laser
pulses. The proposed pulse implements adiabatic rapid passage, which
is a well-established technique for populating discrete states\
\cite{malinovsky01}. It extends the technique, by controlling the pump
spectrum to create an energy-dependent exciton population (Fig.\
\ref{fig:distn}).

The second stage occurs after the pump pulse has passed. It is the
coherent quantum dynamics of the system, starting from the exciton
population created by the pump. The pump is chosen such that this
population is similar to a Fermi distribution, with a sharp upper
step. The system is described by a model similar to that which
describes pair condensation in atomic gases and superconductors. We
therefore expect that a population with the form of a Fermi
distribution could condense, due to a dynamical version of the BCS
instability\ \cite{eastham07}.

We now turn to the theoretical demonstration of this proposal. For
simplicity, we suppose that the pump is circularly polarized, so we may
consider only one of the polarization states of the excitons. We model
the quantum dots as a set of two-level systems, each describing the
presence or absence of an exciton of the pump polarization in a given
localized dot state. These localized excitons are coupled to the
electromagnetic field by the dipole interaction. Since the exciton
states in the dots are spatially separated, we neglect the
non-radiative interactions between the different two-level
systems. The appropriate Hamiltonian is then the generalized Dicke
model\ \cite{KES+05}.

We label the dot states with an index $i$, so that $E_i$ is the energy
of an exciton in the $i^{th}$ dot state; this state is localized at
$\mathbf{r}_i$, with dipole-coupling strength $g_i$. The area density
of dots is $n$, so that if $g_i=g$ and $E_i=E$ the vacuum Rabi
splitting at resonance is $2g\sqrt{n}$. The state of a dot is
specified by the Bloch vector $\langle \bm{\sigma} \rangle$, where
$\sigma^-_i=\sigma^{x}-i\sigma^{y}$ is the exciton annihilation
operator, and $\sigma^z$ the inversion. The inversion is related to
the occupation of the dot $n_i$ by $n_i=(\sigma^z+1)/2$, and is
$-1$(+1) for an unoccupied (occupied) dot.

Dicke models have previously been used to describe polariton
condensation in equilibrium\ \cite{EL01,KES+05,KRK+06}, and in a
dissipative open system\ \cite{marzdecay}. Here we are concerned with
the opposite limit, of timescales short compared with the relaxation
times. The dynamics therefore obeys the Heisenberg equation. Since we
are concerned with condensation phenomena, involving large photon
numbers, we treat the field classically. However, we retain the full
quantum dynamics of the dots, and hence the possibility of an
incoherent population of excitons.  In this approximation, the
Heisenberg equation gives
\begin{subequations}
\label{eq:mbeqns}
\begin{align} i\dot{\psi}_{\mathbf{k},t}&=\omega_{\mathbf{k}} \psi_{\mathbf{k},t} +
\frac{1}{2}\int g P_{\mathbf{k},t}(E,g) dEdg + F_{\mathbf{k},t}, \label{eq:empsi} \\
i\dot{P}_{\mathbf{k},t}(E,g)&= E P_{\mathbf{k},t}(E,g) - 2g
\sum_{\mathbf{k}^\prime} D_{\mathbf{k}-\mathbf{k}^\prime,t}(E,g)
\psi_{\mathbf{k}^\prime,t},
\label{eq:emp} \\
\begin{split} i\dot{D}_{\mathbf{k},t}(E,g) & = g \sum_{\mathbf{k}^\prime} \big(
\psi_{\mathbf{k}^\prime,t}^\ast P_{\mathbf{k}+ \mathbf{k}^\prime,t}(E,g) \\ 
 & \qquad\qquad\qquad + P^\ast_{\mathbf{k}^\prime-\mathbf{k},t}(E,g)
\psi_{\mathbf{k}^\prime,t}\big). \end{split} \label{eq:emd}
\end{align} \end{subequations} $\psi_{\mathbf{k}}$ is the
normal-mode amplitude for an electromagnetic field mode, with in-plane
wavevector $\mathbf{k}$ and energy $\omega_{\mathbf{k}}$, and
$F_{\mathbf{k}}(t)$ is the (classical) pump. $P_{\mathbf{k},t}(E,g)$
and $D_{\mathbf{k},t}(E,g)$ are collective variables describing the
polarizations and inversions of the dots. They are defined as
\begin{align} P_{\mathbf{k}} (E,g)\delta E \delta g & =
\frac{1}{An}\sideset{}{'}\sum_i \langle \sigma_i^- \rangle e^{-i\mathbf{k}.\mathbf{r}_i},
\\ D_{\mathbf{k}}(E,g) \delta E \delta g & = \frac{1}{An}\sideset{}{'}\sum_i \langle
\sigma_i^z\rangle e^{-i\mathbf{k}.\mathbf{r}_i}, \label{eq:ddefn}
\end{align} where the
sums run over the states with $E\to E+\delta E$ and $g\to g+\delta g$.

Eqs. (\ref{eq:empsi}--\ref{eq:emd}) generalize the Maxwell-Bloch
equations\ \cite{staliunas}, to allow for the distribution of energies
and dipole-coupling strengths in the dots. Thus the inversion and
polarization become distribution functions: $D_{0}(E,g)\delta E\delta
g$ is the inversion due to states with energies $E\to E+\delta E$ and
couplings $g\to g+\delta g$.

The fields $P_{\mathbf{k},t}, \psi_{\mathbf{k},t}$, and
$F_{\mathbf{k},t}$ have been normalized such that their square
magnitudes are particle numbers per exciton state. A condensate is
characterized by a macroscopic occupation number, \emph{i.e.}, one
which scales with the size of the system. Thus an exciton-photon
condensate has at least one $P_{\mathbf{k,t}}\sim N^0$ and one
$\psi_{\mathbf{k,t}} \sim N^0$, where $N$ is the total number of dots
in the active region of the sample. In contrast, in a non-condensed
state there is at most of order one particle per mode, and the
$P_{\mathbf{k},t}$ and $\psi_{\mathbf{k},t}$ are all $\lesssim
N^{-1/2}$.

The approximation leading to (\ref{eq:mbeqns}) is the standard
semiclassical approximation, used to treat condensates including
lasers\ \cite{risken70}, superconductors, BCS superfluids\
\cite{andreev04,barankov04}, and polariton condensates\
\cite{eastham05}. It neglects the quantum fluctuations of the
electromagnetic field, which dominate if the photon number is small,
\emph{i.e.}, close to threshold with a small number of dots\
\cite{risken70,eastham05}. Since we could have $N\sim 10^3 $, the
semiclassical approximation is in general very
well-controlled. However, to obtain a correct description of the
dynamics, we must supplement (\ref{eq:mbeqns}) with noise
terms. Without such terms, the non-condensed solution remains a
steady-state above threshold, as in all semiclassical treatments of
condensation. However, it becomes unstable, and hence is not
realized. We therefore add a perturbation driving the field into
(\ref{eq:empsi}), modeling for example spontaneous emission into the
cavity modes. The form and strength of this perturbation does not
affect our results: we show results with Gaussian white noise, but
have obtained similar results using a delta-like kick.

We now specialize (\ref{eq:mbeqns}) to develop a simulation of our
proposed experiment. We introduce imaginary parts to
$\omega_{\mathbf{k}}$ to allow for the decay of the microcavity
photons, with timescales of a few picoseconds\ \cite{DPG+06}. The
initial condition is $\langle \sigma_i^z\rangle=-1$, and there are many
exciton states distributed over the active area of the sample. Thus
the sum in (\ref{eq:ddefn}) is strongly peaked near $\mathbf{k}=0$,
and we approximate the initial conditions as
$D_{\mathbf{k},t}(E,g)=\delta_{\mathbf{k}}D_{0,t}(E,g)$. The dynamics
is then that of a continuous medium due to motional narrowing, with
the short-range spatial structure of the exciton states averaged out
on the long scales of the photons.

We consider a plane-wave pump, at a high angle where the excitons lie
outside the stop-bands of the mirrors. The field acting on the dots at
this wavevector, $\mathbf{k}_p$, may then be taken as the driving
field.  Anticipating our analysis of the condensation, we retain only
one other mode of the field, specifically the confined cavity mode
with $\mathbf{k}=0$. This reduces \eqref{eq:mbeqns} to
\begin{subequations} \label{eq:twomodemb}
\begin{align}
i\dot\psi_0&=\omega_0 \psi_0 + \frac{1}{2}\int g P_0(E,g) dEdg,
\label{eq:em2modepsi0} \\ i\dot P_0(E,g)&=EP_0(E,g)-2gD_0(E,g)\psi_0, \label{eq:em2modep0} \\ i\dot
P_p(E,g)&=EP_p(E,g)-2gD_0(E,g)F_p, \label{eq:em2modepp} \\ 
\begin{split} i\dot D_0(E,g)&=g \left(F_p^\ast
P_p(E,g)+P_p^\ast(E,g) F_p \right. \\ &  \left. \quad +\psi_0^\ast P_0(E,g)+P_0^\ast(E,g) \psi_0\right). \end{split},  \label{eq:em2moded0}
\end{align} \end{subequations} corresponding to an ensemble of two-level
systems, interacting with two modes of the field. $P_p$ is the polarization at
the pump wavevector, and $F_p$ the driving field. 

\begin{figure}[t]
\includegraphics[width=246pt]{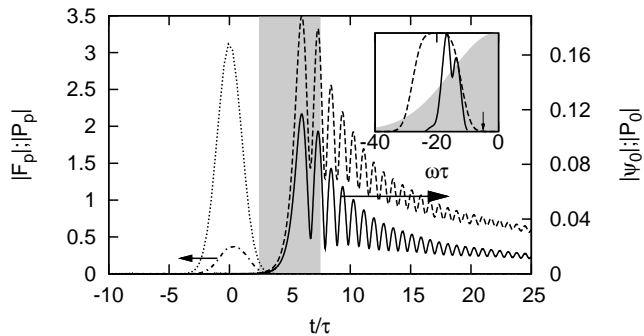}
\caption{Electromagnetic fields and polarizations as functions of
  time, for the simulation described in the text. Dotted: pump field
  $|F_{p}|$ (left axis). Dot-dashed: polarization $|P_p|$ at pump
  wavevector, integrated over dot energies and couplings (left
  axis). Solid: cavity field $|\psi_0|$ at $\mathbf{k}=0$ (right
  axis). Dashed: polarization $|P_0|$ at $\mathbf{k}=0$ (right
  axis). Inset: spectrum $|\psi_0(\omega)|^2$ during the shaded region
  of the main plot (solid curve). The pumped population (dashed
  curve), exciton energy distribution (shading), and energy of the
  $\mathbf{k}=0$ cavity mode (arrow) are shown for
  comparison. \label{fig:fieldsim}}
\end{figure}

\begin{figure}[t]
\includegraphics[width=246pt]{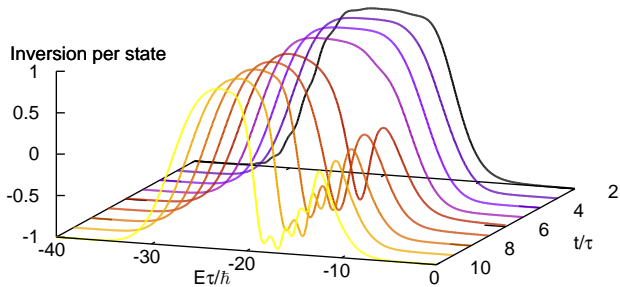}
\caption{(Color online) Simulated exciton inversion profile during the
condensation shown in Fig.\ \ref{fig:fieldsim}. \label{fig:pevolve}}
\end{figure}

We have simulated our proposed experiment by solving
(\ref{eq:twomodemb}) numerically, with $N=4500$ two-level
systems. Results are shown in Figs.\
\ref{fig:fieldsim}--\ref{fig:pevolve}. These results focus on a model
with a single coupling strength $g$, and a Gaussian distribution of
exciton energies with variance $\sigma^2$. The pump is a
linearly-chirped Gaussian,
\begin{equation}2 g F_p(t)=\frac{S}{\sqrt{2\pi\tau^2}}e^{-i(\nu_0+\alpha
t/2)t} e^{-t/(2\tau^2)}, \label{eq:gausspulse}
\end{equation} where $S=\int 2g |\psi(t)|
dt$ is the usual pulse area per exciton state. The pulse time $\tau$
defines dimensionless times and energies, and the zero of energy is
chosen at the center of the exciton line. The remaining parameters are
$\sigma=15\hbar/\tau$, $\Re(\omega_0)=-5\hbar/\tau$,
$\nu_0=-20\hbar/\tau$, $\alpha=5/\tau^2$, $-\Im(\omega_{0})=1.5/\tau$,
$g=13\bar/\tau$ and $S=5\pi$. These parameters, with $\tau=3\;
\mathrm{ps}$, are reasonable for a microcavity containing interfacial
quantum dots. Though there is a distribution of $g$, due to the
different sizes of the dot states, this does not qualitatively change
our results.

Figs.\ \ref{fig:fieldsim}--\ref{fig:pevolve} are the key results of
this paper, demonstrating the scenario outlined in our
introduction. Referring first to Fig.\ \ref{fig:fieldsim}, we see that
there can indeed be two stages to the dynamics. In the first stage,
during the pump pulse, $P_0$ and $\psi_0$ are vanishingly small. $P_p$
does become finite, reflecting the fact that the pump laser does
induce some coherent polarization in the excitons. This coherence is
small -- $P_p \approx 1/\sqrt{N}\approx 0.02$ -- but more importantly
both $P_p$ and $F_p$ have disappeared by the end of the pumping
stage. We therefore argue that the pump produces an incoherent
population of excitons. Furthermore, as shown in Fig.\
\ref{fig:distn}, this population has a sharp upper step, like that of
the Fermi function at a low effective temperature $T\sim 1/\tau$.

In the second stage, visible in Fig.\ \ref{fig:fieldsim}, we see both
$\psi_0$ and $P_0$ building up to values of order $N^0$. This directly
demonstrates condensation of excitons and photons.

To understand the condensation, we consider the dynamics of the pumped
population. Linearizing (\ref{eq:mbeqns}) gives normal modes
$\psi_{\mathbf{k},t}=e^{i\lambda_{\mathbf{k},t}}A_\psi$,
$P_{\mathbf{k},t}=e^{i\lambda_{\mathbf{k},t}}A_P(E,g)$. Their
frequencies
$\lambda_{\mathbf{k}}=\lambda_{\mathbf{k}}^\prime+i\lambda_{\mathbf{k}}^{\prime\prime}$
obey\ \cite{eastham07}
\begin{equation}\omega_{\mathbf{k}}-\lambda_{\mathbf{k}} = - \int
  \frac{\nu(E)}{E-\lambda_{\mathbf{k}}}
  dE,\label{eq:stabeqn}\end{equation} where $\nu(E)=\int g^2 D_0(E,g)
dg$ is an optical density of the excitons. Applying this result to
the state immediately after the pump pulse, we find an unstable mode
at $\mathbf{k}=0$. This instability gives the exponential growth of
the polarization and field, to macroscopic values, visible in Fig.\
\ref{fig:fieldsim}. In fact, (\ref{eq:stabeqn}) predicts
instabilities for $|\mathbf{k}|<k_c$. The fastest-growing instability is at 
$\mathbf{k}$=0, and as this mode grows, it suppresses the gain for the others. It will thus be dynamically selected, and we therefore neglected cavity modes with
$\mathbf{k}\neq 0$ in the simulations.

For the parameters used here, the instability predicted by
(\ref{eq:stabeqn}) corresponds to the BCS instability in a
superconductor. This can be seen by considering (\ref{eq:stabeqn}) for
a single coupling strength, close to an instability. The eigenenergies
$\lambda_{\mathbf{k}}^\prime$ and growth rates
$\lambda_{\mathbf{k}}^{\prime\prime}$ then obey\begin{subequations}
  \label{eq:linstab} \begin{align}
    \frac{\omega^\prime_{\mathbf{k}}-\lambda_{\mathbf{k}}^\prime}{g^2}
    &= - \mathcal{P}\int \frac{D_0(E)}{E-\lambda_{\mathbf{k}}^\prime}
    dE,
  \label{eq:instfreq} \\
\lambda_{\mathbf{k}}^{\prime\prime}&=\pi g^2\mathrm{sgn}
(\lambda_{\mathbf{k}}^{\prime\prime})
D_0(\lambda_{\mathbf{k}}^\prime)-\gamma. \label{eq:gainloss}
\end{align} \end{subequations} (\ref{eq:gainloss}) describes the growth or decay of the normal mode,
with the first term the gain/loss from the excitons, and the second
the loss due to the cavity decay
$\gamma=-\Im({\omega_{\mathbf{k}}})$. (\ref{eq:instfreq}) is the
Cooper equation of the BCS model\ \cite{degennes}. The term
corresponding to the usual pairing interaction is
$g^2/(\lambda_{\mathbf{k}}-\omega_{\mathbf{k}})$, which we recognize
as the effective interaction between excitons, mediated by the cavity
modes. Here it is an attractive interaction, as required for BCS,
since the excitons lie below the photons. The term corresponding to
the Fermi distribution is the exciton population created by the
pump. In a superconductor, there is a solution to the Cooper equation
below the Fermi energy, due to the step in the Fermi distribution; in
the same way, (\ref{eq:instfreq}) has a solution below the step in the
exciton occupation. (\ref{eq:gainloss}) shows that this mode
experiences gain, and hence can become unstable.

To confirm the origin of the condensation, we show in the inset to
Fig.\ \ref{fig:fieldsim} the spectrum of the $\mathbf{k}=0$ field, and
in Fig.\ \ref{fig:pevolve} the evolution of the population during
condensation. As expected, the condensate is at a frequency below the
step in the exciton population (Fig.\ \ref{fig:fieldsim}); this leads
to hole-burning there (Fig.\ \ref{fig:pevolve}). Reducing the center
frequency of the chirp, $\nu_0$, we find that the condensation stage
in Fig.\ \ref{fig:fieldsim} disappears, as the effective pairing
interaction decreases below its critical value. The long-time limit of
the simulations is then just the population of excitons. We have also
confirmed that the phase of the pump (\ref{eq:gausspulse}) is
irrelevant, by plotting the phase associated with the power spectrum
in the inset to Fig.\ \ref{fig:fieldsim}. This plot is identical in
simulations with different pump phases, but the same random noise.

Since our condensate occurs on timescales short compared with the
relaxation times, it will not be in equilibrium. This leads to
phenomena absent for an equilibrium condensate. In Fig.\
\ref{fig:fieldsim}, for example, we see ringing oscillations
(corresponding to those predicted for atomic gases\
\cite{andreev04,barankov04,yuzbashyan06,eastham07}; unrelated
oscillations have been predicted in coherently-driven microcavities\
\cite{shelykhspind05}), and a slow decay. These phenomena will be
discussed in a future publication.

It is perhaps surprising that our pump configuration could generate an
incoherent population, without macroscopic occupation. To explain the
mechanism, we note first that the driving field dominates during the
pump stage (Fig.\ \ref{fig:fieldsim}). Thus the pumping can be
understood in terms of the well-known dynamics of non-interacting
two-level systems, driven by a chirped laser pulse\
\cite{malinovsky01}. After eliminating the time-dependent frequency
$\omega(t)$ of the pump with a unitary transformation, this dynamics
is a precession of the Bloch vectors $\langle \bm{\sigma}_i\rangle$
around axes $\mathbf{B}_i=(2 g_i R(t),0,\omega(t)-E_i)$, where $R(t)$
is the pump field at the dot.

Consider first a dot with energy inside the chirp. Provided the pump
is strong enough, the Bloch vector of such a dot adiabatically follows
$\mathbf{B}_i$ from $-\mathbf{\hat k}$ to $\mathbf{\hat k}$, and it
becomes populated. Furthermore, a dot with energy outside the range of
the chirp will not respond unless the pump is very strong, and hence
such a dot will remain unpopulated. Thus we see that the only dots
which could be polarized by the pump are those at the edge of the
chirp. However, there are very few such dots. Furthermore, any
polarizations induced in dots with different energies would have
different phases, due to the chirping.  Thus, as shown in Fig.\
\ref{fig:fieldsim}, the chirped pump does not induce a collective
polarization. Noting additionally that there are no cavity modes
resonant with the pump, we see why the pump generates a state with
$P_{\mathbf{k},t}, \psi_{\mathbf{k},t} \lesssim N^{-1/2}$.

A fully quantitative description of our proposed experiment is likely
to require a more detailed model of the quantum dot states. It would
also be desirable to extend our model to (a)incorporate relaxation
processes, which on longer timescales will drive the off-equilibrium
condensate towards an equilibrium one\ \cite{eastham07}, and
(b)develop a full quantum theory, as has been done for photon lasers\
\cite{risken70}, polariton lasers\ \cite{krizhan2,laussy:016402}, and
equilibrium condensates\ \cite{eastham05}. Finally, it would be
interesting to compare this work with Ref.\ \
\cite{vinck-posada:167405}, which appears to show a type of dynamical
condensation, under very different conditions.

Though the parameters used here have been chosen to suit interfacial
quantum dots\ \cite{SP02,LP05}, many other choices are possible. The
basic requirements are that the pumping is slow enough to create a
controlled inversion profile within the inhomogeneous line, such that
the couplings can be increased so that the instability occurs after
the pump pulse, but before the excitons decay. Other systems described
by generalized Dicke models, such as Fermi gases, SK dots in
microcavities, or Josephson junction arrays, could be considered.

To conclude, we have proposed and analyzed a new approach to quantum
condensation in a solid-state system. The key is the chirped pumping
(\ref{eq:gausspulse}), which we have shown can create an
energy-dependent population in the exciton line. Such a population can
condense, even in the absence of relaxation or inelastic
scattering. Since our approach uses the spectrum of the pump to tailor
the population, it would be possible to pump other initial states
(inset to Fig.\ \ref{fig:distn}). Thus our technique could be used
more generally, to explore the quantum dynamics of a many-particle
system from controlled initial conditions.

This work was supported by EPSRC grants EP/F040075/1 and EP/C546814/01.


\end{document}